\documentclass[floatfix,rmp,twocolumn,twoside]{revtex4h}
\usepackage{graphicx}

\bibpunct{[}{]}{,}{n}{}{,}

\usepackage[english]{babel}
\usepackage{verbatim}
\usepackage{subfigure}
\usepackage{multirow}
\usepackage{varwidth}
\usepackage{graphicx}

\newcommand{\etexttt}[1]{\textit{#1}}
\newcommand{\an}[1]{\emph{#1}} 

\newenvironment{smallverbatim}
{\footnotesize\verbatim}{\endverbatim}

\newcommand{\sepv}{\vspace{-6.0mm}}

\begin{document}

\title{A Tool for Model-Based Language Specification}
\author{Luis~Quesada, Fernando~Berzal, and Juan-Carlos~Cubero\\
  Department of Computer Science and Artificial Intelligence, CITIC, University of Granada, \\
  Granada 18071, Spain \\
  \textit{lquesada@decsai.ugr.es, fberzal@decsai.ugr.es, jc.cubero@decsai.ugr.es}
  }

\begin{abstract}

%
%
Formal languages let us define the textual representation of data with precision. Formal grammars, typically in the form of BNF-like
productions, describe the language syntax, which is then annotated for syntax-directed translation and completed with semantic actions.
%
%
When, apart from the textual representation of data, an explicit representation of the corresponding data structure is required, the language
designer has to devise the mapping between the suitable data model and its proper language specification, and then develop the conversion
procedure from the parse tree to the data model instance.
%
%
Unfortunately, whenever the format of the textual representation has to be modified, changes have to propagated throughout the entire language
processor tool chain. These updates are time-consuming, tedious, and error-prone.
%
%
Besides, in case different applications use the same language, several copies of the same language specification have to be maintained.
%
%
In this paper, we introduce a model-based parser generator that decouples language specification from language processing, hence avoiding many
of the problems caused by grammar-driven parsers and parser generators.
%

\end{abstract}

\maketitle

\section{Introduction}


%
%
A formal language represents a set of strings \cite{Jurafsky2009}.
%
%
Formal languages consist of an alphabet, which describes the basic symbol or character set of the language, and a grammar, which describes how
to write valid sentences of the language.
%
%
In Computer Science, formal languages are used, among other things, for the precise definition of data formats and the syntax of programming
languages.
%
%
The front-end of a language processor, such as an interpreter or compiler, determines the grammatical structure corresponding to the textual
representation of data conforming to a given language specification. Such grammatical structure is typically represented in the form of a parse
tree.
%

%
%
Most existing language specification techniques \cite{Aho1972} require the language designer to provide a textual specification of the language
grammar.
%
%
The proper specification of such a grammar is a nontrivial process that depends on the lexical and syntactic analysis techniques to be used,
since each kind of technique requires the grammar to comply with different constraints. Each analysis technique is characterized by its
expression power and this expression power determines whether a given analysis technique is suitable for a particular language. The most
significant constraints on formal language specification originate from the need to consider context-sensitivity, the need of performing an
efficient analysis, and some techniques' inability to resolve conflicts caused by grammar ambiguities.
%


%
%
In its most general sense, a model is anything used in any way to represent something else. In such sense, a grammar is a model of the language
it defines.
%
%
In Software Engineering, data models are also common. Data models explicitly determine the structure of data. Roughly speaking, they describe
the elements they represent and the relationships existing among them.
%
%
From a formal point of view, it should be noted that data models and grammar-based language specifications are not equivalent, even though both
of them can be used to represent data structures. A data model can express relationships a grammar-based language specification cannot.
Moreover, a data model does not need to comply with the constraints a grammar-based language specification has to comply with. Hence describing
a data model is generally easier than describing the corresponding grammar-based language specification.
%

%
%
When both a data model and a grammar-based language processor are required, the implementation of the language processor requires the language
designer to build a grammar-based language specification from the data model and also to implement the conversion from the parse tree to the
data model instance.
%

%
%
Whenever the language specification has to be modified, the language designer has to manually propagate changes throughout the entire language
processor tool chain, from the specification of the grammar defining the formal language (and its adaptation to specific parsing tools) to the
corresponding data model. These updates are time-consuming, tedious, and error-prone. By making such changes labor-intensive, the traditional
approach hampers the maintainability and evolution of the language \cite{Kats2010}.
%

%
%
Moreover, it is not uncommon that different applications use the same language. For example, the compiler, different code generators, and other
tools within an IDE, such as the editor or the debugger, typically need to grapple with the full syntax of a programming language. Their
maintenance typically requires keeping several copies of the same language specification in sync.
%

%
%
As an alternative approach, a language can also be defined by a data model that, in conjunction with the declarative specification of some
constraints, can be automatically converted into a grammar-based language specification \cite{Quesada2011c}.

%
%
This way, the data model representing the language can be modified as needed without having to worry about the language processor and the
peculiarities of the chosen parsing technique, since the corresponding language processor with be automatically updated.

%
%
Furthermore, as the data model is the direct representation of a data structure, such data structure can be implemented as an abstract data type
(in object-oriented languages, as a set of collaborating classes). Following the proper software design principles, that implementation can be
performed without having to embed or mix semantic actions with the language specification, as it is typically the case with grammar-driven
language processors.

%
%
Finally, as the data model is not bound to a particular parsing technique, evaluating alternative and/or complementary parsing techniques is
possible without having to propagate their constraints into the language model. Therefore, by using an annotated data model, model-based
language specification completely decouples language specification from language processing, which can be performed using whichever parsing
techniques that are suitable for the formal language implicitly defined by the model.
%

%
%
In this paper, we introduce ModelCC, a model-based tool for language specification. As a parser generator that decouples language specification
from language processing, it avoids many of the problems caused by grammar-driven parsers and parser generators.
%

%
%
Section \ref{sec:background} introduces formal grammars and surveys parsing algorithms and tools.
%
%
Section \ref{sec:modelbased} presents the philosophy behind model-based language specification.
%
%
Section \ref{sec:modelspecification} comments on ModelCC building blocks.
%
%
Section \ref{sec:modelconstraints} describes the model constraints supported by ModelCC, which declaratively define the features of the formal
language defined by the model.
%
%
Section \ref{sec:example} shows a prototypical example, which is used to discuss the advantages of model-based language specification over
traditional grammar-based language specification.
%
%
Lastly, Section \ref{sec:conclusionsfuturework} presents our conclusions and the future work that derives from our research.
%

\section{Background} \label{sec:background}

In this section, we introduce formal grammars (Subsection A), describe the typical architecture of language processor front-ends (Subsection B),
survey key parsing algorithms (Subsection C), and review existing lexer and parser generators (Subsection D).

\subsection{Formal Grammars} \label{subsec:formalgrammars}

%
%
Formal grammars are used to specify the syntax of a language \cite{Ginsburg1975,Harrison1978}. A grammar naturally describes the hierarchical
structure of most language constructs \cite{Aho2006}. Using a set of rules, a grammar describes how to form strings from the language alphabet
that are valid according to the language syntax.
A grammar $G$ is formally defined \cite{Chomsky1956} as the tuple $(N,\Sigma,P,S)$, where:
\begin{itemize}
\item $N$ is the finite set of nonterminal symbols of the language, sometimes called syntactic variables, none of which appear in the language strings.
\item $\Sigma$ is the finite set of terminal symbols of the language, also called tokens, which constitute the language alphabet
(i.e. they appear in the language strings). Therefore, $\Sigma$ is disjoint from $N$.
\item
$P$ is a finite set of productions, each one of the form $(\Sigma \cup N)^{*} N (\Sigma \cup N)^{*} \rightarrow (\Sigma \cup N)^{*}$, where $*$
is the Kleene star operator, $\cup$ denotes set union, the part before the arrow is called the left-hand side of the production, and the part
after the arrow is called the right-hand side of the production.
\item $S$ is a distinguished nonterminal symbol, $S \in N$: the grammar start symbol.
\end{itemize}
%

%
%
%
For convenience, when several productions share their left-hand side, they can be grouped into a single production containing their the shared
left-hand side and all their different right-hand sides separated by $|$.
%

%
%
Context-free grammars are formal grammars in which the left-hand side of each production consists of only a single nonterminal symbol. All their
productions, therefore, are of the form $N \rightarrow (\Sigma \cup N)^{*}$. Context-free grammars generate context-free languages.
%

%
%
A context-free grammar is said to be ambiguous if there exists at least a string that can be generated by the grammar in more than one way. In
fact, some context-free languages are inherently ambiguous (i.e. all context-free grammars generating them are ambiguous).
%

%
%
An attribute grammar is a formal way to define attributes for the symbols in the productions of a formal grammar, associating values to these
attributes. Semantic rules annotate attribute grammar and define the value of an attribute in terms of the values of other attributes and
constants \cite{Aho2006}. They key feature of attribute grammars is that they let us transfer information from anywhere in the abstract syntax
tree to anywhere else in a controlled and formal way, hence their frequent use in syntax-directed translation.

%

%
%
Graph grammars \cite{Ehrig1999} allow the manipulation of graphs based on productions: if the left-hand side of a production matches the working
graph or a subgraph of it, it can be replaced with the right-hand side of the production. These grammars can be used to define the syntax of
visual languages.
%

\subsection{The Architecture of Language Processors}

%
%
The architecture of a language-processing system decomposes language processing into several steps which are typically grouped into two phases:
analysis and synthesis. The analysis phase, which is responsibility of the language processor front end, starts by breaking up its input into
its constituent pieces (lexical analysis or scanning) and imposing a grammatical structure upon them (syntax analysis or parsing). The language
processor back end will later synthesize the desired target from the results provided by the front end.
%

%
%
A lexical analyzer, also called lexer or scanner, processes an input string conforming to a language specification and produces the tokens found
within it.
%
%
Lexical ambiguities occur when a given input string simultaneously corresponds to several token sequences \cite{Nawrocki1991}, which may
overlap.
%

%
%
A syntactic analyzer, also called parser, processes input tokens and determines their grammatical structure with respect to the given language
grammar, usually in the form of parse trees. In the absence of lexical ambiguities, the parser input consists of a stream of tokens, whereas it
will be a directed acyclic graph of tokens when lexical ambiguities are present.
%
%
Syntactic ambiguities occur when a given set of tokens simultaneously corresponds to several parse trees \cite{Aho1975}.
%

\vspace{-3mm} 

\subsection{Scanning and Parsing Algorithms}

%
%
Scanning and parsing algorithms are characterized by the expression power of the languages they can apply to, their support for ambiguities or
lack thereof, and the constraints they impose on language specifications.
%

%
%
Traditional lexers are based on a finite-state machine that is built from a set of regular expressions \cite{McNaughton1960}, each of which
describes a token type. The efficiency of regular expression lexers is $O(n)$, being $n$ the input string length.
%

%
%
Lamb \cite{Quesada2011a} is a lexer with lexical ambiguity support that allows the specification of tokens not only by regular
expressions, but also by arbitrary pattern matching classes. Lamb also supports token type precedences.
The efficiency of the Lamb lexer is, in the worst case, $O(n^2t^2)$, being $n$ the input string length and $t$ the number of different
token types.
%

%
%
Efficient parsers for specific subsets of context-free grammars exist. These include top-down LL parsers, which construct a leftmost derivation
of the input sentence, and bottom-up LR parsers, which construct a rightmost derivation of the input sentence.

LL grammars were formally introduced in \cite{Lewis1968}, albeit LL(k) parsers predate their name \cite{Oettinger1961}. An LL parser is called
an LL(k) parser if it uses $k$ tokens of lookahead when parsing a sentence, while it is an LL(*) parser if it is not restricted to a finite $k$
tokens of lookahead and it can make parsing decisions by recognizing whether the following tokens belong to a regular language
\cite{Jarzabek1975,Nijholt1976}. While LL(k) parsers are always linear, LL(*) ranges from $O(n)$ to $O(n^2)$.

LR parsers were introduced by Knuth \cite{Knuth1965}. DeRemer later developed the LALR \cite{DeRemer1969,DeRemer1982} and SLR
\cite{DeRemer1971} parsers that are in use today. When parsing theory was originally developed, machine resources were scarce, and so parser
efficiency was the paramount concern \cite{Parr2011}. Hence all the aforementioned parsing algorithms parse in linear time (i.e. their
efficiency is $O(n)$, being $n$ the input string length) and they do not support syntactic ambiguities.

Efficient LR and LL parsers for certain classes of ambiguous grammars are also possible by using simple disambiguating rules \cite{Aho1975,Earley1975}.

%
%
A general-purpose dynamic programming algorithm for parsing context-free grammars was independently developed by Cocke \cite{Cocke1970}, Younger
\cite{Younger1967}, and Kasami \cite{Kasami1965}: the CYK parser. This general-purpose algorithm is $O(n^3)$ for ambiguous and unambiguous
context-free grammars. The Earley parser \cite{Earley1970} is another general-purpose dynamic programming algorithm for parsing context-free
grammars that executes in cubic time ($O(n^3)$) in the general case, quadratic time ($O(n^2)$) for unambiguous grammars, and linear time
($O(n)$) for almost all LR(k) grammars.


Free from the requirement to develop efficient linear-time parsing algorithms, researchers have developed many powerful nondeterministic parsing
strategies following both the top-down approach (LL parsers) and the bottom-up approach (LR parsers).

%
%
Following the top-down approach, Packrat parsers \cite{Ford2002packrat} and their associated Parsing Expression Grammars (PEGs)
\cite{Ford2004peg} preclude only the use of left-recursive grammar rules and force the rules to be ordered, that is, when an alternative succeeds, the following alternatives are ignored.
Even though they use backtracking, packrat parsers are linear rather than exponential because of the rule order and because they memoize partial results.

%
%
Following the bottom-up approach, Generalized LR (GLR) parsers parse in linear to cubic time, depending on how closely the grammar conforms to
the underlying LR strategy. The time required to run the algorithm is proportional to the degree of nondeterminism in the grammar.
Bernard Lang is typically credited with the original GLR idea \cite{Lang1974}. Later, Tomita used the algorithm for natural language processing
\cite{Tomita1985}. Tomita's Universal parser \cite{Tomita1987}, however, failed for grammars with epsilon rules (i.e. productions with an
empty right-hand side). Several extensions have been proposed that support epsilon rules \cite{Farshi1991,Rekers1992,Ishii1994,McPeak2004}.

%

GLR parsing is an optimization of Earley's algorithm and, like parsing expression grammars, tends to be scannerless. Being scannerless is
necessary if a tool needs to support grammar composition (i.e. to easily integrate one language into another or create new grammars by modifying
and composing pieces from existing grammars), unless lexical ambiguities are supported.
The Fence parser \cite{Quesada2011b} is an
optimized Earley-like algorithm that works on top of the Lamb lexer \cite{Quesada2011a}, which provides support for lexical ambiguities. Hence,
Fence is also suitable for grammar composition and as efficient as GLR parsers in practice.
%

\vspace{-6mm} 

\subsection{Lexer and Parser Generators}

%
%
Lexer and parser generators are tools that take a language specification as input and produce a lexer or parser as output. They can be
characterized by their input syntax, their ability to specify semantic actions, and the parsing algorithms the resulting parsers implement.
%

%
%
Lex \cite{lex} and yacc \cite{yacc} are commonly used in conjunction \cite{Levine1992}. They are the default lexer generator and
parser generator in many Unix environments and standard compiler textbooks use them as examples, e.g. \cite{Aho2006}. Lex is the
prototypical regular-expression-based lexer generator, while yacc and its derivatives generate LALR parsers.
%

%
%
JavaCC \cite{McManis1996} is a parser generator that creates LL(k) parsers, albeit it has been superseded by ANTLR \cite{Parr1995}.
ANTLR is a scannerless parser generator that creates LL(*) parsers. ANTLR-generated parsers are linear in practice and greatly reduce
speculation, reducing the memoization overhead of pure packrat parsers.

%

The Rats! \cite{Grimm2006} packrat parser generator is a PEG-based tool that also optimizes memoization to improve its speed and reduce its
memory footprint. Like ANTLR, it does not accept left-recursive grammars. Unlike ANTLR, programmers do not have to deal with conflict messages,
since PEGs have no concept of a grammar conflict: they always choose the first possible interpretation, which can lead to unexpected behavior.

%
%
NLyacc \cite{Ishii1994} and Elkhound \cite{McPeak2004} are examples of GLR parser generators. Elkhound achieves yacc-like
parsing speeds when grammars are LALR(1) but suffers from the practical limitations of GLR parsers. Like PEG parsers, GLR parsers do not always
do what is intended, in part because they accept ambiguous grammars and programmers have to detect ambiguities dynamically
\cite{Parr2011}.

%

%
%
YAJco \cite{Poruban2009} is an interesting tool that accepts, as input, a set of Java classes with annotations that specify the prefixes,
suffixes, operators, tokens, parentheses, and optional elements common in typical programming languages. As output, YAJco generates a
BNF-like grammar specification for JavaCC \cite{McManis1996}. Since YAJco is built on top of a parser generator, the language designer
has to be careful when annotating his classes, as the implicit grammar he is defining has to comply with the constraints imposed by the
underlying LL(k) parser generator.
%

%
%
To the best of our knowledge, no existing parser generator follows the approach we now proceed to describe.
%

\section{Model-Based Language Specification} \label{sec:modelbased}

In this section, we discuss the concepts of abstract and concrete syntax, analyze the potential advantages of model-based language
specification, and compare our proposed approach with the traditional grammar-driven language design process.

\subsection{Abstract Syntax and Concrete Syntaxes}

%
%
The abstract syntax of a language is just a representation of the structure of the different elements of a language without the superfluous
details related to its particular textual representation \cite{Kleppe2007}.
%
%
On the other hand, a concrete syntax is a particularization of the abstract syntax that defines, with precision, a specific textual or graphical
representation of the language.
%
%
It should also be noted that a single abstract syntax can be shared by several concrete syntaxes \cite{Kleppe2007}.
%

%
%
For example, the abstract syntax of the typical \emph{$<$if$>$-$<$then$>$-$<$optional else$>$} statement in imperative programming languages
could be described as the concatenation of a conditional expression and one or two statements. Different concrete syntaxes could be defined for
such an abstract syntax, which would correspond to different textual representations of a conditional statement, e.g. \{``if'', ``('',
expression, ``)'', statement, optional ``else'' followed by another statement\} and \{``IF'', expression, ``THEN'', statement, optional ``ELSE''
followed by another statement, ``ENDIF''\}.
%

The idea behind mode-based language specification is that, starting from a single abstract syntax model (ASM) representing the core concepts in
a language, language designers would later develop one or several concrete syntax models (CSMs). These concrete syntax models would suit the
specific needs of the desired textual or graphical representation. The ASM-CSM mapping could be performed, for instance, by annotating the
abstract syntax model with the constraints needed to transform the elements in the abstract syntax into their concrete representation.

\subsection{Advantages of Model-Based Language Specification}

%
%
Focusing on the abstract syntax of a language offers some benefits \cite{Kleppe2007} and provides some potential advantages to model-based
language specification over the traditional grammar-based language specification approach:

\begin{itemize}

\item
When reasoning about the features a language should include, specifying its abstract syntax seems to be a better starting point than working on
its concrete syntax details. In fact, we control complexity by building abstractions that hide details when appropriate \cite{sicp}.

\item
Sometimes, different incarnations of the same abstract syntax might be better suited for different purposes (e.g. an human-friendly syntax for
manual coding, a machine-oriented format for automatic code generation, a Fit-like \cite{fit} syntax for testing, different architectural views
for discussions with project stakeholders...). Therefore, it might be useful for a given language to support multiple syntaxes.

\item
Since model-based language specification is independent from specific lexical and syntactic analysis techniques, the constraints imposed by
specific parsing algorithms do not affect the language design process. In principle, it might not be even necessary for the language designer to
have advanced knowledge on parser generators when following a model-driven language specification approach.

\item
A full-blown model-driven language workbench \cite{language-workbenches} would allow the modification of a language abstract syntax model and
the automatic generation of a working IDE on the run. The specification of domain-specific languages would become easier, as the language
designer could play with the language specification and obtain a fully-functioning language processor on the fly, without having to worry about
the propagation of changes throughout the complete language processor tool chain.
\end{itemize}

In short, the model-driven language specification approach brings domain-driven design \cite{ddd} to the domain of language design. It provides
the necessary infrastructure for what Evans would call the `supple design' of language processing tools: the intention-revealing specification
of languages by means of abstract syntax models, the separation of concerns in the design of language processing tools by means of declarative
ASM-CSM mappings, and the automation of a significant part of the language processor implementation.

\subsection{Comparison with the Traditional Approach}

\begin{figure}[tb!]
\centering
\includegraphics[scale=0.24]{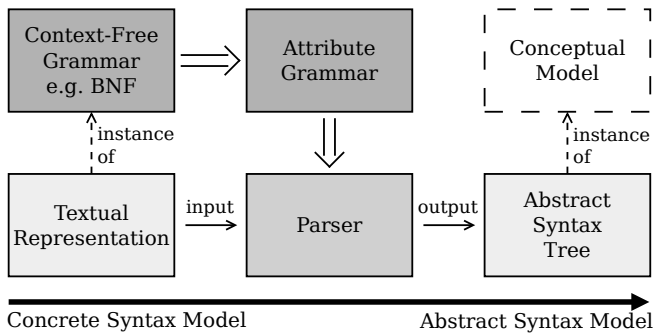}
\caption{Traditional language processing approach.} \label{fig:traditional}
\end{figure}

\begin{figure}[tb!]
\centering
\includegraphics[scale=0.24]{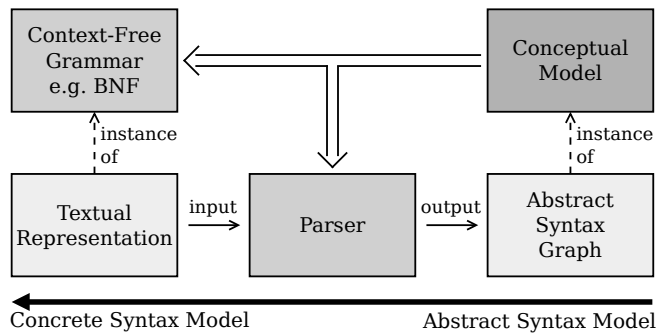}
\caption{Model-based language processing approach.} \label{fig:ModelCC}
\end{figure}

A diagram summarizing the traditional language design process is shown in Figure \ref{fig:traditional}, whereas the corresponding diagram for
the model-based approach proposed in this paper is shown in Figure \ref{fig:ModelCC}.

When following the traditional grammar-driven approach, the language designer starts by designing the grammar corresponding to the concrete
syntax of the desired language, typically in BNF or a similar format. Then, the designer annotates the grammar with attributes and, probably,
semantic actions, so that the resulting attribute grammar can be fed into lexer and parser generator tools that produce the corresponding lexer
and parser, respectively. The resulting syntax-directed translation process generates abstract syntax trees from the textual representation in
the concrete syntax of the language.


When following the model-driven approach, the language designer starts by designing the conceptual model that represents the abstract syntax of
the desired language, focusing on the elements the language will represent and their relationships. Instead of dealing with the syntactic
details of the language from the start, the designer devises a conceptual model for it (i.e. the abstract syntax model, or ASM), the same way a
database designer starts with an implementation-independent conceptual database schema before he converts that schema into a logical schema that
can be implemented in the particular kind of DBMS that will host the final database. In the model-driven language design process, the ASM would
play the role of entity-relationship diagrams in database design and each particular CSM would correspond to the final table layout of the
physical database schema in a relational DBMS.

Even though the abstract syntax model of the language could be converted into a suitable concrete syntax model automatically, the language
designer will often be interested in specifying the details of the ASM-CSM mapping. With the help of constraints imposed over the abstract
model, the designer will be able to guide the conversion from the ASM to its concrete representation using a particular CSM. This concrete
model, when it corresponds to a textual representation of the abstract model, will be described by a formal grammar. It should be noted,
however, that the specification of the ASM is independent from the peculiarities of the desired CSM, as a database designer does not consider
foreign keys when designing the conceptual schema of a database. Therefore, the grammar specification constraints enforced by particular parsing
tools will not impose limits on the design of the ASM. The model-driven language processing tool will take charge of that and, ideally, it will
employ the most efficient parsing technique that works for the language resulting from the ASM-CSM mapping.


While the traditional language designer specifies the grammar for the concrete syntax of the language, annotates it for syntax-directed
processing, and obtains an abstract syntax tree that is an instance of the implicit conceptual model defined by the grammar, the model-based
language designer starts with an explicit full-fledged conceptual model and specifies the necessary constraints for the ASM-CSM mapping. In both
cases, parser generators create the tools that parse the input text in its concrete syntax. The difference lies in the specification of the
grammar that drives the parsing process, which is hand-crafted in the traditional approach and automatically-generated as a result of the
ASM-CSM mapping in the model-driven process.


Another difference stems from the fact that the result of the parsing process in an instance of an implicit model in the grammar-driven approach
while that model is explicit in the model-driven approach. An explicit conceptual model is absent in the traditional language design process
albeit that does not mean that it does not exist. On the other hand, the model-driven approach enforces the existence of an explicit conceptual
model, which lets the proposed approach reap the benefits of domain-driven design.


There is a third difference between the grammar-driven and the model-driven approaches to language specification. While, in general, the result
of the parsing process is an abstract syntax tree that corresponds to a valid parsing of the input text according to the language concrete
syntax, nothing prevents the conceptual model designer from modeling non-tree structures. Hence the use of the `abstract syntax graph' term in
Figure \ref{fig:ModelCC}. This might be useful, for instance, for modeling graphical languages, which are not constrained by the linear nature
of the traditional syntax-driven specification of text-based languages.


Instead of going from a concrete syntax model to an implicit abstract syntax model, as it is typically done, the model-based language
specification process goes from the abstract to the concrete. This alternative approach facilitates the proper design and implementation of
language processing systems by decoupling language processing from language specification, which is now performed by imposing declarative
constraints on the ASM-CSM mapping.


\section{ModelCC Model Specification} \label{sec:modelspecification}

Once we have described model-driven language specification in general terms, we now proceed to introduce ModelCC, a tool that supports our
proposed approach to the design of language processing systems. ModelCC, at its core, acts as a parser generator. Its starting abstract syntax
model is created by defining classes that represent language elements and establishing relationships among those elements (associations in UML
terms). Once the abstract syntax model is established, its incarnation as a concrete syntax is guided by the constraints imposed over language
elements and their relationships as annotations on the abstract syntax model. In other words, the declarative specification of constraints over
the ASM establishes the desired ASM-CSM mapping.

In this section, we introduce the basic constructs that allow the specification of abstract syntax models, while we will discuss how model
constraints help us establish a particular ASM-CSM mapping in the following section of this paper. Basically, the ASM is built on top of basic
language elements, which might be viewed as the tokens in the model-driven specification of a language. ModelCC provides the necessary
mechanisms to combine those basic elements into more complex language constructs, which correspond to the use of concatenation, selection, and
repetition in the syntax-driven specification of languages.

Our final goal is to allow the specification of languages in the form of abstract syntax models such as the one shown in Figure
\ref{fig:calcmodelcc}, which will be used as an example in Section \ref{sec:example}. This model, in UML format, specifies the abstract syntax
model of the language supported by a simple arithmetic calculator. The annotations that accompany the model provide the necessary information
for establishing the complete ASM-CSM mapping that corresponds to the traditional infix notation for arithmetic expressions. Moreover, the model
also incorporates the method that lets us evaluate such arithmetic expressions. Therefore, Figure \ref{fig:calcmodelcc} represents a complete
interpreter for arithmetic expressions in infix notation using ModelCC (its complete implementation as a set of cooperating Java classes appears
in Figure \ref{fig:calcimmodelcc}).

As mentioned above, the specification of the ASM in ModelCC starts with the definition of basic language elements, which can be modeled as
simple classes in an object-oriented programming language. The ASM-CSM mapping of those basic elements will establish their correspondence to
the tokens that appear in the concrete syntax of the language whose ASM we design in ModelCC. In the following subsections, we describe the
mechanisms provided by ModelCC to implement the three main constructs that let us specify complete abstract syntax models on top of basic
language elements.

\subsection{Concatenation}

Concatenation is the most basic construct we can use to combine sets of language elements into more complex language elements. In textual
languages, this is achieved just by joining the strings representing its constituent language elements into a longer string that represents the
composite language element.

In ModelCC, concatenation is achieved by object composition. The resulting language element is the composite element and its members are the
language elements the composite element collates.

When translating the ASM into a textual CSM, each composite element in a ModelCC model generates a production rule in the grammar representing
the CSM. This production, with the nonterminal symbol of the composite element in its left-hand side, concatenates the nonterminal symbols
corresponding to the constituent elements of the composite element in its right-hand side. By default, the order of the constituent elements in
the production rule is given by the order in which they are specified in the object composition, but such an order is not strictly necessary
(e.g. many ambiguous languages might require unordered sequences of constituent elements and even some unambiguous languages allow for such
flexibility).

\begin{figure}[tb!]
\centering
\includegraphics[scale=1]{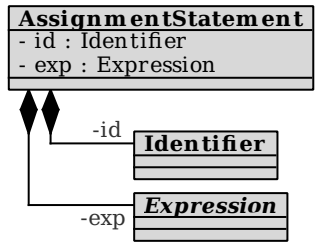}
\caption{An assignment statement as an example of element composition (concatenation in textual CSM terms).} \label{fig:assignmentstatement}
\end{figure}

The model in Figure \ref{fig:assignmentstatement} shows an example of object composition in ASM terms that corresponds to string concatenation
in CSM terms. In this example, an assignment statement is composed of an identifier, i.e. a reference to its l-value, and an expression, which
provides its r-value. In a textual CSM, the composite \emph{AssignmentStatement} element would be translated into the following production rule:
\etexttt{$<$AssignmentStatement$>$ ::= $<$Identifier$>$ $<$Expression$>$}. Obviously, such production would probably include some syntactic
sugar in an actual programming language, either for avoiding potential ambiguities or just for improving its readability and writability, but
that is the responsibility of ASM-CSM mappings, which will be analyzed in Section \ref{sec:modelconstraints}.

\subsection{Selection}

Selection is necessary as a language modeling primitive operation to represent choices, so that we can specify alternative elements in language
constructs.

In ModelCC, selection is achieved by subtyping. Specifying inheritance relationships among language elements in an object-oriented context
corresponds to defining `is-a' relationships in a more traditional database design setting. The language element we wish to establish
alternatives for is the superelement (i.e. the superclass in OO or the supertype in DB modeling), whereas the different alternatives are
represented as subelements (i.e. subclasses in OO, subtypes in DB modeling). Alternative elements are always kept separate to enhance the
modularity of ModelCC abstract syntax models and their integration in language processing systems.

In the current version of ModelCC, multiple inheritance is not supported, albeit the same results can be easily simulated by combining
inheritance and composition. We can define subelements for the different inheritance hierarchies representing choices so that those subelements
are composed by the single element that appears as a common choice in the different scenarios. This solution fits well with most existing
programming languages, which do not always support multiple inheritance, and avoids the pollution of the shared element interface in the ASM,
which would appear as a side effect of allowing multiple inheritance in abstract syntax models.

Each inheritance relationship in ModelCC, when converting the ASM into a textual CSM, generates a production rule in the CSM grammar. In those
productions, the nonterminal symbol corresponding to the superelement appears in its left-hand side, while the nonterminal symbol of the
subelement appears as the only symbol in the production right-hand side. Obviously, if a given superelement has $k$ different subelements, $k$
different productions will be generated representing the $k$ alternatives defined by the abstract syntax model.

\begin{figure}[tb!]
\centering
\includegraphics[scale=1]{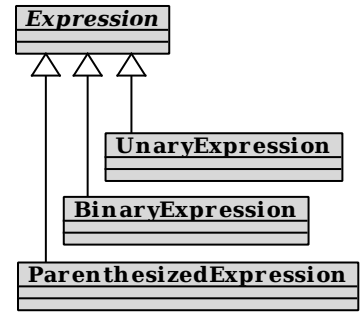}
\caption{Subtyping for representing choices in ModelCC.} \label{fig:expression}
\end{figure}

For example, the model shown in Figure \ref{fig:expression} illustrates how an arithmetic \emph{Expression} can be an \emph{UnaryExpression}, a
\emph{BinaryExpression}, or a \emph{ParenthesizedExpression} in the language defined for a simple calculator. The grammar resulting from the
conversion of this ASM into a textual CSM would be: \etexttt{$<$Expression$>$ ::= $<$UnaryExpression$>$ $|$ $<$BinaryExpression$>$ $|$
$<$ParenthesizedExpression$>$}.

\subsection{Repetition}

Representing repetition is also necessary in abstract syntax models, since a language element might appear several times in a given language
construct. When a variable number of repetitions is allowed, mere concatenation does not suffice.

Repetition is also achieved though object composition in ModelCC, just by allowing different multiplicities in the associations that connect
composite elements to their constituent elements. The cardinality constraints described in Section \ref{sec:cardinality} can be used to annotate
ModelCC models in order to establish specific multiplicities for repeatable language elements.

Each composition relationship representing a repetitive structure in the ASM will lead to two additional production rules in the grammar
defining a textual CSM: a recursive production of the form \etexttt{$<$ElementList$>$ ::= $<$Element$>$ $<$ElementList$>$} and a complementary
production \etexttt{$<$ElementList$>$ ::= $<$Element$>$}, where \etexttt{$<$Element$>$} is the nonterminal symbol associated to the repeating
element. Obviously, an equivalent non-left-recursive derivation can also be obtained when needed.

It should also be noted that \etexttt{$<$ElementList$>$} will take the place of the nonterminal \etexttt{$<$Element$>$} in the production
derived from the composition relationship that connects the repeating element with its composite element (see the above section on how
composition is employed to represent concatenation in ModelCC).

In practice, repeating elements will often appear separated in the concrete syntax of a textual language, hence repeatable elements can be
annotated with separators, as we will see in Section \ref{sec:delimiters}. In case separators are employed, the recursive production derived
from repeatable elements will be of the form \etexttt{$<$ElementList$>$ ::= $<$Element$>$ $<$Separator$>$ $<$ElementList$>$}.

When a repeatable language element is optional, i.e. its multiplicity can be $0$, an additional epsilon production has to be appended to the
grammar defining the textual CSM derived from the ASM: \etexttt{$<$ElementList$>$ ::= $\epsilon$}.

\begin{figure}[tb!]
\centering
\includegraphics[scale=1]{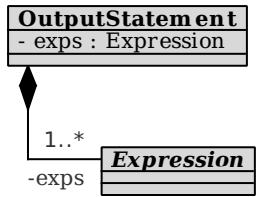}
\caption{Multiple composition for representing repetition in ModelCC.} \label{fig:output}
\end{figure}

For example, the model in Figure \ref{fig:output} shows that an \emph{OutputStatement} can include several \emph{Expression}s, which will be
evaluated for their results to be sent to the corresponding output stream. This ASM would result in the following textual CSM grammar:
\etexttt{$<$OutputStatement$>$ ::= $<$ExpressionList$>$} for describing the composition and \etexttt{$<$ExpressionList$>$ ::= $<$Expression$>$
$<$ExpressionList$>$ $|$ $<$Expression$>$} for allowing repetition.

\section{ModelCC Model Constraints} \label{sec:modelconstraints}

Once we have examined the mechanisms that let us create abstract syntax models in ModelCC, we now proceed to describe how constraints can be
imposed on such models in order to establish the desired ASM-CSM mapping. As soon as that ASM-CSM mapping is established, ModelCC is able to
generate the suitable parser for the concrete syntax defined by the CSM.


In ModelCC, the constraints imposed over abstract syntax models to define a particular ASM-CSM mapping are declared as metadata annotations on
the model itself. Now supported by all the major programming platforms, metadata annotations have been used in reflective programming and code
generation \cite{Fowler2002}. Among many other things, they can be employed for dynamically extending the features of your software development
runtime \cite{Berzal2005} or even for building complete model-driven software development tools that benefit from the infrastructure provided by
your compiler and its associated tools \cite{mdsd-ideal}.


In ModelCC, which supports language composition without being scannerless, metadata annotations are used for pattern specification, a necessary
feature for defining the lexical elements of the concrete syntax model, i.e. its tokens (Subsection A).

Annotations are also employed for defining delimiters in the concrete syntax model, whose use is common for eliminating language ambiguities or
just as syntactic sugar in many languages (Subsection B).

A third group of ModelCC metadata annotations lets us impose cardinality constraints on language elements, which control element repeatability
and optionality (Subsection C).

Finally, a fourth set of metadata annotations lets us impose evaluation order constraints in ModelCC, which are employed to declaratively
resolve further ambiguities in the concrete syntax of a textual language, by establishing associativity, precedence, and composition
constraints, the latter employed for resolving the ambiguities that cause the typical shift-reduce conflicts in LR parsers (Subsection D).

A summary of the complete set of annotations supported by ModelCC can be found at the end of this section, after the detailed description of
each of the four groups of ModelCC metadata annotations.

\vspace{-3mm} 

\subsection{Pattern Specification Constraints}


Pattern specification constraints allow the specification of the lexical elements in a concrete syntax model, i.e. the different token types
defined for the concrete syntax of a textual language. It should be noted that, once a language element is annotated with pattern specification
constraints, it cannot be a composite element since, as a lexical element, it cannot be composed of other elements.


The above constraint, which forces lexical elements to be basic language elements in a ModelCC ASM, does not reduce the flexibility of ModelCC
for language composition. Language composition, typically achieved by scannerless parser generators, can also be achieved if the scanner
supports lexical ambiguities. When lexical ambiguities are allowed, as in the Lamb scanning algorithm \cite{Quesada2011a}, the same string or
even overlapping strings might return several tokens, which will later be processed by the parser consuming the output of the scanner supporting
lexical ambiguities.

\vspace{-3mm} 
\subsubsection{The @Pattern annotation}


The \an{@Pattern} annotation allows the specification of the pattern that will be used to match a basic language element in the input string.
Two mutually exclusive mechanisms are provided for pattern specification in ModelCC: regular expressions and user-defined pattern matching
classes. Regular expressions can be specified in ModelCC to build standard lexers, whereas custom pattern matching classes allow the language
designer to use any custom-defined matching element to recognize basic language elements in the input string. The custom pattern matching class
can be anything, since it works as a black box for ModelCC. It might even be a complete ModelCC-generated parser, which could be used for the
specification of modular languages, a coarse form of language composition (e.g. think of the JavaScript scripts and CSS stylesheets within the
HTML in a web page).



When used with regular expressions, the \an{@Pattern} annotation includes an argument representing the regular expression. This regular
expression, specified as the \emph{regExp} field of the annotation, corresponds to the traditional token type definition in lex-like scanners.


When used with custom pattern matching classes, the \an{@Pattern} annotation is used to specify the name of the class implementing the matching
algorithm and its argument string. In this case, ModelCC resorts to the Lamb lexer \cite{Quesada2011a}, which will employ the
pattern matching class specified by the \emph{matcher} field of the \an{@Pattern} annotation to process the input string. This
Lamb-specific \cite{Quesada2011a} lexical definition makes use of Lamb support for lexical ambiguities, overlapping tokens, and
black-box token recognizers.

\begin{figure}[tb!]
\centering
\includegraphics[scale=1]{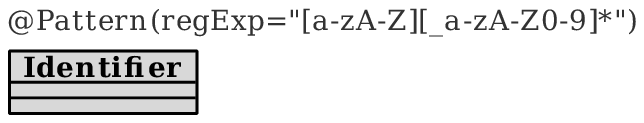}
\caption{Pattern specification example: Regular expression.} \label{fig:identifier}
\end{figure}

For example, the \an{@Pattern} annotation in Figure \ref{fig:identifier} defines the typical \emph{Identifier} token in programming languages,
which can be specified by the following regular expression: \emph{[a-zA-Z][\_a-zA-Z0-9]*}. This specification corresponds to the lex
token definition shown in Figure \ref{fig:identifierg}.

\begin{figure}[tb!]
\begin{smallverbatim}

[a-zA-Z][_a-zA-Z0-9]*  return Identifier;
\end{smallverbatim}
\caption{Implementation of Figure \ref{fig:identifier} in lex.}
\label{fig:identifierg}
\end{figure}


\begin{figure}[tb!]
\centering
\includegraphics[scale=1]{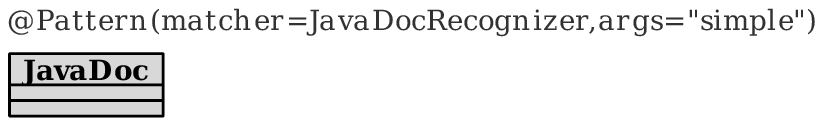}
\caption{Pattern specification example: Custom pattern matching.} \label{fig:javadoc}
\end{figure}

The example in Figure \ref{fig:javadoc} illustrates the use of custom pattern matching classes in ModelCC. In this case, the \an{@Pattern}
annotation refers to the \emph{JavaDocRecognizer} class, which will be responsible for recognizing \emph{JavaDoc} comments as basic language
elements in the shown ASM. Arguments can also be specified when using the \emph{matcher} field of the \an{@Pattern} annotation with the help of
its optional \emph{args} field (\emph{``simple''} in Figure \ref{fig:javadoc}).

As mentioned above, the ModelCC use of custom pattern matching algorithms for defining basic language elements has no counterpart in traditional
lexer generators, hence an equivalent lex-like definition cannot be provided. In ModelCC, the Lamb scanning algorithm \cite{Quesada2011a} will be responsible for processing
such token definitions.

\subsubsection{The @Value annotation}


The \an{@Value} annotation can be used in combination with the \an{@Pattern} annotation in ModelCC to indicate the location where the recognized
token value will be stored in the abstract syntax graph, so that value can be used once the input string has been parsed.

\begin{figure}[tb!]
\centering
\includegraphics[scale=1]{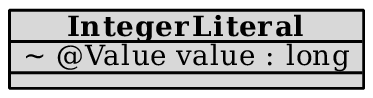}
\caption{Value field specification example: Integer literals.}
 \label{fig:integerliteral}
\end{figure}

\begin{figure}[tb!]
\begin{smallverbatim}

[0-9]+ {
  yylval.value = atoi(yytext);
  return INTEGERLITERAL;
}
\end{smallverbatim}
\caption{Implementation of Figure \ref{fig:integerliteral} using lex \& yacc.} \label{fig:integerliteralg}
\end{figure}


Associated to a field of the class defining a basic language element, that field will contain the value obtained from the input string that
matches the token type pattern specification.

When a numeric or boolean field is annotated with the \an{@Value} annotation, it is not necessary to specify the corresponding \an{@Pattern}
annotation for recognizing the numeric or boolean tokens. When the \an{@Value}-annotated field is not numeric nor boolean (e.g. a string, a
single character, or any non-primitive data type), the use of the \an{@Pattern} annotation is mandatory.


Elements from the ASM that contain a \an{@Value}-annotated numeric or boolean field are transformed into their corresponding token type lex-like
definitions, even when no \an{@Pattern} annotation is present. ModelCC will always perform the proper assignment of the recognized token to the
\an{@Value}-annotated field.


For example, the model in Figure \ref{fig:integerliteral} defines an \emph{IntegerLiteral} language element that recognizes long integer
literals. The proper regular expression for such literals will be employed and, whenever an integer literal is found in the input string, its
integer value will be stored in the \an{@Value}-annotated \emph{value} field of the \emph{IntegerLiteral} class. If we had used lex \&
yacc, we would have had to type the lexical definition and semantic action shown in Figure \ref{fig:identifierg}.

\begin{figure}[tb!]
\centering
\includegraphics[scale=1]{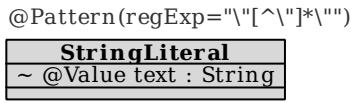}
\caption{Value field specification example: Double-quoted string literals.} \label{fig:stringliteral}
\end{figure}

As another example, Figure \ref{fig:stringliteral} shows how the \an{@Value} annotation would be used in conjunction with the \an{@Pattern}
annotation to define string literals surrounded by double quotes. A \emph{StringLiteral} will be recognized whenever a pair of double quotes
encloses a string not containing a double quote, a constraint that can be specified by the following regular expression:
\emph{$\backslash$``[$\mathbin{\char`\^}\backslash$``]*$\backslash$``}. The lex \& yacc implementation of this string literal
token type definition would be slightly more complex than the previous example, as shown in Figure \ref{fig:stringliteralg}.

\begin{figure}[tb!]
\begin{smallverbatim}

\"[^\"]*\" {
  int size,csize;
  size = strlen(yytext)-2;
  csize = sizeof(char)*(size+1);
  yylval.pval = malloc(csize);
  strncpy(yylval.pval,yytext+1,size);
  yylval.pval[size] = '\0';
  return STRINGLITERAL;
}
\end{smallverbatim}
\caption{Implementation of Figure \ref{fig:stringliteral} using lex \& yacc.} \label{fig:stringliteralg}
\end{figure}


\subsection{Delimiter Constraints}
\label{sec:delimiters}


Delimiter constraints allow the specification of language element delimiters in a concrete syntax model. Delimiters include prefixes, suffixes,
and separators. Such kinds of delimiters are often used to eliminate language ambiguities and facilitate parsing, but they can appear just as
syntactic sugar to make languages more readable.


Usually, reserved words in programming languages act just as delimiters. As such, they will not appear in the language abstract syntax model.
They will be specified as metadata annotations in the ASM-CSM mapping corresponding to the concrete syntax of the language.


It should be noted that delimiters should always be specified as constraints on the ASM-CSM mapping rather than language elements in the ASM,
since they do not provide any relevant information from the perspective of the abstract syntax model, even though they might be necessary to
define unambiguous textual languages.

\subsubsection{The @Prefix annotation}


The \an{@Prefix} annotation allows the specification of prefixes for language elements and specific constituents in composite language elements.


The \emph{value} field of the \an{@Prefix} annotation is used to specify the list of regular expressions that define the prefixes that precede
the corresponding language element (or a specific constituent element within a composite element) in the concrete syntax of a textual CSM.

When converting the ASM into a textual CSM, ModelCC will include the specified prefixes in every production where the annotated element appears
in the textual CSM grammar, just before the appearance of the annotated element. When the annotation is associated to a constituent element
within a composite language element, the sequence of prefixes will be included only in the productions that correspond to the composite language
element, preceding the annotated constituent element within their right-hand side.

It should be noted that, when the annotated element is repeatable, the sequence of prefixes appear only once, preceding the first instance of
the annotated element. Prefixes will also be included in the CSM even when no elements appear in a repetition language construct (e.g. as when
the opening parenthesis appears before an empty list of arguments in a parameterless C-like function call).

\begin{figure}[tb!]
\centering
\includegraphics[scale=1]{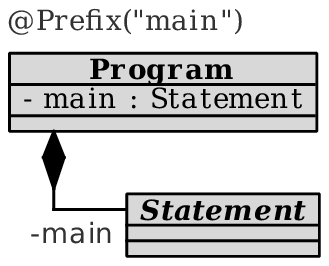}
\caption{\an{@Prefix} annotation example.} \label{fig:programmain}
\end{figure}

For example, the model in Figure \ref{fig:programmain} specifies that the textual representation of a \emph{Program} will always be preceded by
a ``main'' keyword prefix. The grammar defining the textual CSM for this simple example is  \etexttt{$<$ProgramMain$>$ ::= ``main''
$<$Statement$>$}.

\subsubsection{The @Suffix annotation}


The \an{@Suffix} annotation allows the specification of suffixes for language elements and specific constituents in composite language elements.


The \emph{value} field of the \an{@Suffix} annotation is used to specify the list of regular expressions that define the suffixes that follow
the corresponding language element (or a specific constituent element within a composite element) in the concrete syntax of a textual CSM.

When converting the ASM into a textual CSM, ModelCC will include the specified suffixes in every production where the annotated element appears
in the textual CSM grammar, just after the appearance of the annotated element. When the annotation is associated to a constituent element
within a composite language element, the sequence of suffixes will be included only in the productions that correspond to the composite language
element, following the annotated constituent element within their right-hand side.

It should be noted that, when the annotated element is repeatable, the sequence of suffixes appear only once, just after the last element in the
sequence of repetitions. Suffixes will also be included in the CSM even when no elements appear in a repetition language construct (e.g. as when
the closing parenthesis appears at the end of an empty list of arguments in a parameterless C-like function call).

\begin{figure}[tb!]
\centering
\includegraphics[scale=1]{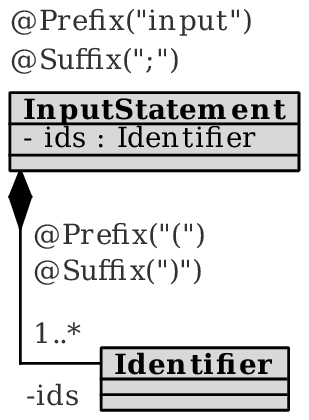}
\caption{\an{@Suffix} annotation example.} \label{fig:inputstatement}
\end{figure}

For example, the model in Figure \ref{fig:inputstatement} specifies that the textual representation of an \emph{InputStatement} is preceded by an
``input'' keyword prefix and followed by a semicolon (``;'') as its suffix. It also contains a sequence of \emph{Identifier}s delimited by
opening and closing parentheses: ``('' as the prefix of the \emph{ids} constituent of the \emph{InputStatement} composite and ``)'' as its
suffix. The grammar defined by the ASM-CSM mapping specified by the annotations in Figure \ref{fig:inputstatement} is
\etexttt{$<$InputStatement$>$ ::= ``input'' ``('' $<$IdentifierList$>$ ``)'' ``';''; $<$IdentifierList$>$ ::= $<$Identifier$>$ $<$IdentifierList$>$ $|$
$<$Identifier$>$}.

\subsubsection{The @Separator annotation}


The \an{@Separator} annotation allows the specification of separators between consecutive instances of elements within a repetition. Separators
can be defined in ModelCC by annotating a language element in the ASM or just its appearance within a particular repetition construct. In the
first case, a default separator is established for the language element: the specified separator will be used for separating consecutive
instances of the annotated language element whenever a sequence of such language elements appears in a textual CSM. In the second case, an
\emph{ad hoc} separator is defined: the specified separator will be used only when consecutive instances of the language element appear within
the context of the annotated repetition construct.


The \emph{ad hoc} definition of separators with the \an{@Separator} annotation within repetition constructs can be used to override the default
sequence of separators associated to the repeatable element in a repetition construct (or just to disable the use of separators for that
specific repetition).

Therefore, default separators are specified for language elements by the \an{@Separator} annotation. When converting the ASM into a textual CSM,
ModelCC will include the sequence of regular expressions defining those default separators in every recursive production rule generated from a
repetition where the annotated element is repeatable. When the \an{@Separator} annotation accompanies a repeatable element within a particular
repetition construct, i.e. the \emph{ad hoc} case, separators will only appear in the recursive production rule derived from that particular
repetition construct, but not in other constructs where the constituent element might also be repeatable.

\begin{figure}[tb!]
\centering
\includegraphics[scale=1]{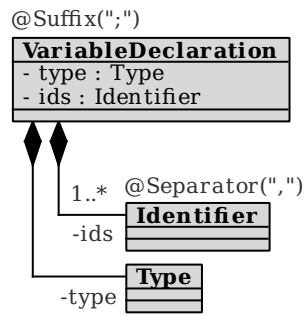}
\caption{Default \an{@Separator} example.} \label{fig:identifier2}
\end{figure}

As an example of defining a default separator, Figure \ref{fig:identifier2} illustrates how a comma (``,'') can be used as the default separator
for \emph{Identifier}s. Whenever a list of \emph{Identifier}s is needed within the language, a comma will separate consecutive identifiers. In
the example, since a \emph{VariableDeclaration} contains a \emph{Type} and a set of \emph{Identifier}s, they will be separated by ``,'' in the
textual CSM derived from the language ASM. The grammar of the resulting CSM will include the following productions:
\etexttt{$<$VariableDeclaration$>$ ::= $<$Type$>$ $<$IdentifierList$>$ ``;''} and \etexttt{$<$IdentifierList$>$ ::= $<$Identifier$>$ ``,''
$<$IdentifierList$>$ $|$ $<$Identifier$>$}.

\begin{figure}[tb!]
\centering
\includegraphics[scale=1]{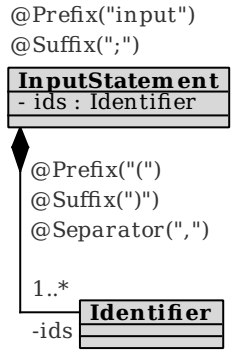}
\caption{\emph{Ad hoc} \an{@Separator} example.} \label{fig:inputstatement2}
\end{figure}

As an example illustrating the use of \emph{ad hoc} separators, consider the model in Figure \ref{fig:inputstatement2}. Here, \emph{Identifier}s
are also separated by commas, but only within \emph{InputStatement}s, i.e. ``,'' is the \emph{ad hoc} separator for identifiers within input
statements, but lists of identifiers might employ different separators elsewhere. The grammar associated to the textual CSM derived from Figure
\ref{fig:inputstatement2} would include the following productions: \etexttt{$<$InputStatement$>$ ::= ``input'' ``(''
$<$InputStatementIdentifierList$>$ ``)'' ``;''} and \etexttt{$<$InputStatementIdentifierList$>$ ::= $<$Identifier$>$ ``,''
$<$InputStatementIdentifierList$>$ $|$ $<$Identifier$>$}.

\vspace{-3mm} 

\subsection{Cardinality Constraints}
\label{sec:cardinality}


A third group of ModelCC metadata annotations lets us impose cardinality constraints on language elements, which control element repeatability
and optionality.

\vspace{-3mm} 
\subsubsection{The @Optional annotation}


The \an{@Optional} annotation allows the specification of optional elements in textual CSMs.

Optional elements naturally appear in language specifications and optionality could always be modeled by means of selection constructs. However,
the declarative specification of the optionality constraints is necessary to avoid unnecessary duplication in the language model.

When one of the constituent elements within a composite language element is optional, the textual representation of the composite element might
include the optional element, along with its corresponding delimiters, or not. In the latter case, the missing element delimiters are not
included in the textual representation the composite element either, even though a prefix and a suffix might have been defined for the missing
constituent element.



If we performed a naive transformation of a composite language into a set of CFG production rules and the composite element includes $i$
optional elements, $2^i$ production rules would result with the composite element in their left-hand side and all the possible combinations of
optional elements in their right-hand side. A more reasonable transformation employs just $i$ ancillary epsilon production rules.

\begin{figure}[tb!]
\centering
\includegraphics[scale=1]{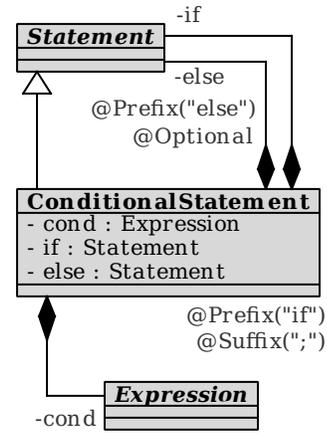}
\caption{\an{@Optional} element example: if-then-else statement.} \label{fig:conditional2}
\end{figure}

For example, the model in Figure \ref{fig:conditional2} shows that a \emph{ConditionalStatement} contains an \emph{Expression}, the
\emph{Statement} that will be run when the \emph{Expression} evaluates to true, and, optionally, the \emph{Statement} that will be run when the
\emph{Expression} evaluates to false. The grammar resulting from the model transformation into a textual CSM will include the following two
productions: \etexttt{$<$ConditionalStatement$>$ ::= ``if'' $<$Expression$>$ $<$Statement$>$ $<$OptionalElse$>$} and \etexttt{$<$OptionalElse$>$
::= ``else'' $<$Statement$>$ $|$ $\epsilon$}.

\subsubsection{The @Minimum annotation}


The \an{@Minimum} annotation, depicted as a minimum multiplicity constraint in standard UML notation, allows the specification of the lower
bound for the multiplicity of repeatable language elements within repetition constructs. This lower bound is $1$ by default.

It should be noted that, when the minimum multiplicity is $0$, no elements might appear in a particular instance of the repetition. However,
delimiters would still be represented in the textual CSM unless the \an{@Optional} annotation were explicitly employed.


ModelCC generates semantic actions that check that multiplicity constraints are satisfied whenever they are specified in the model. The
Fence parser \cite{Quesada2011b} allows semantic actions to implement such multiplicity checks and, when they are not satisfied, the
corresponding reduction is automatically inhibited.

Other parser generators would require the explicit generation of a grammar representing the minimum cardinality constraint, i.e. when an element
must appear at least $i$ times within a repetition, two production rules would be necessary: \etexttt{$<$MinIElements$>$ ::= $<$Element$>$... i
times... $<$Element$>$ $<$ElementList$>$} and \etexttt{$<$ElementList$>$ ::= $<$Element$>$ $<$ElementList$>$ $|$ $\epsilon$}.

\begin{figure}[tb!]
\centering
\includegraphics[scale=1]{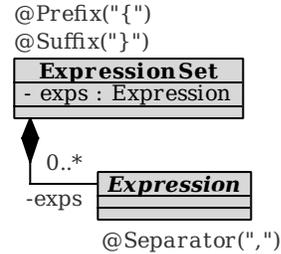}
\caption{Minimum multiplicity example.} \label{fig:expressionset}
\end{figure}

For example, the model in Figure \ref{fig:expressionset} specifies that an \emph{ExpressionSet}, which might include $0$ or more \emph{Expression}s.
In UML notation, no explicit \an{@Minimum} annotation is needed, since the minimum multiplicity constraint in the \emph{exps} association has
the same purpose. The grammar corresponding to the textual CSM derived from Figure \ref{fig:expressionset} would include the following productions
to represent the possibility of empty sets: \etexttt{$<$ExpressionSet$>$ ::= ``\{'' $<$OptionalExpressionList$>$ ``\}'';
$<$OptionalExpressionList$>$ ::= $<$ExpressionList$>$ $|$ $\epsilon$; $<$ExpressionList$>$ ::= $<$Expression$>$ ``,'' $<$ExpressionList$>$ $|$
$<$Expression$>$}.

\subsubsection{The @Maximum annotation}


The \an{@Maximum} annotation, depicted as a maximum multiplicity constraint in standard UML associations, allows the specification of the upper
bound for the multiplicity of repeatable language elements within repetition constructs. This upper bound is undefined by default.


ModelCC incorporates semantic actions that check that multiplicity constraints are satisfied whenever they are specified in the model. The
Fence parser \cite{Quesada2011b} allows semantic actions to implement such multiplicity checks so that, when they are not satisfied, the
corresponding reduction is automatically inhibited. In other words, if an specific upper bound is surpassed for a list of elements, the
generated parser would not recognize the list of elements as such, since it does not satisfy the cardinality constraint imposed by the maximum
multiplicity annotation.

Other parser generators would require the generation of more complex grammars to support maximum multiplicity constraints. In general, if $i$ is
the maximum multiplicity, $i$ alternative production rules would be necessary (assuming that the default minimum multiplicity holds, i.e. $1$).
When both the minimum $i$ and the maximum $j$ multiplicities are specified, one production would be used for representing the minimum
multiplicity constraint and $j-i$ additional productions would be needed to enforce the maximum multiplicity constraint. More complex
combinations of multiplicity constraints could also be devised.

\begin{figure}[tb!]
\centering
\includegraphics[scale=1]{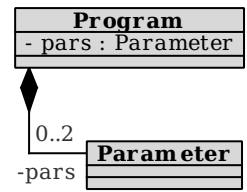}
\caption{Maximum multiplicity example.} \label{fig:arguments}
\end{figure}

For example, the model in Figure \ref{fig:arguments} indicates that a \emph{Program} might have from $0$ to $2$ \emph{Parameter}s. Here, both a
minimum and a maximum multiplicity annotations are inferred from the multiplicity of the UML association. The grammar corresponding to the
textual CSM derived from Figure \ref{fig:arguments} would include the following production rules: \etexttt{$<$Program$>$ ::=
$<$OptionalParameterList$>$; $<$OptionalParameterList$>$ ::= $<$ParameterList$>$ $|$ $\epsilon$; $<$ParameterList$>$ ::= $<$Parameter$>$
$<$ParameterList$>$ $|$ $<$Parameter$>$}, where the last production would be accompanied by a semantic action that would check whether the
maximum multiplicity constraint holds. If such feature were not available in our parsing algorithm generator, this last production would have to
be replaced by a much more explicit (and potentially verbose) set of equivalent productions incorporating the maximum multiplicity constraint:
\etexttt{$<$ParameterList$>$ ::= $<$Parameter$>$ $|$ $<$Parameter$>$ $<$Parameter$>$}. This approach poses no problems in this simple example,
but it might get much more complicated (no problem yet whenever the resulting grammar is automatically generated by a model-driven language
specification tool).

\subsection{Evaluation Order Constraints}
\label{sec:evaluation-order}


A fourth set of ModelCC metadata annotations lets us impose evaluation order constraints, which are employed to declaratively resolve syntactic
ambiguities in the concrete syntax of a textual language by establishing associativity, composition, and precedence constraints for CSMs.

\subsubsection{The @Associativity annotation}

The \an{@Associativity} annotation allows the specification of the operator-like associativity of language elements. ModelCC supports the
following options for specifying associativity constraints:
\begin{itemize}
%
%
\item \an{UNDEFINED}, when no associativity is declared (by default), all possibilities are considered.
%
%
\item \an{LEFT\_TO\_RIGHT}, for left-associative operations (e.g. substraction, division, or function application).
%
%
\item \an{RIGHT\_TO\_LEFT}, for right-associative elements (e.g. exponentiation and function definition).
%
%
\item \an{NON\_ASSOCIATIVE}, for non-associative elements (e.g. cross of three vectors).
\end{itemize}


The specification of associativity constrains help us resolve ambiguities that might appear in recursive compositions (i.e. when using the
composite design pattern \cite{gof} for modeling the ASM for operations without explicit delimiters), where different interpretations of the
input string could be given unless the associativity constraints impose an order on the reductions that can be performed (either left-to-right
or right-to-left).


For each production in the CSM grammar where an the nonterminal of a language element with associativity constraints is preceded and/or followed
by the nonterminal that appears on the left-hand side the production or any of its superclasses in the ASM, ModelCC generates a semantic action
enforcing that associativity constraint. That semantic action, which inhibits the reduction of the production when the constraint is not met, is
directly supported by the Fence parsing algorithm \cite{Quesada2011b}. Fence inhibits the corresponding reduction in three
situations:

\begin{itemize}
\item Whenever the element that follows a left-to-right associative element was generated by a reduction of the same production.
\item Whenever the element that precedes a right-to-left associative element was generated by a reduction of the same production.
\item Whenever an element that precedes or follows a non-associative element was generated by a reduction of the same production.
\end{itemize}

The parsing algorithms implemented by other parser generators also offer mechanisms to enforce associativity constraints.

\begin{figure}[tb!]
\centering
\includegraphics[scale=1]{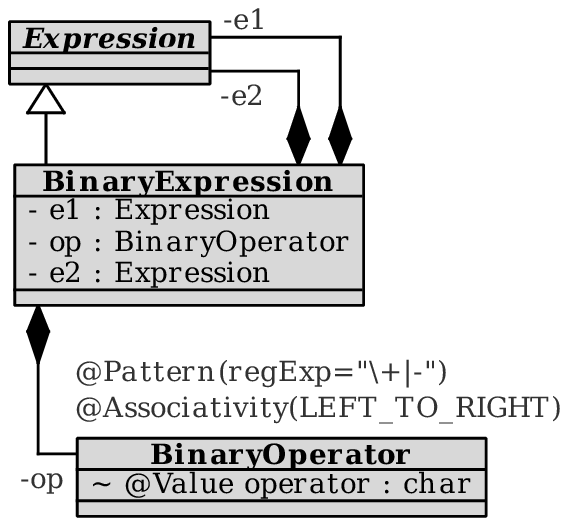}
\caption{Associativity constraint example.} \label{fig:operator}
\end{figure}

For example, the model in Figure \ref{fig:operator} establishes that \emph{BinaryOperator}s are left-associative. The grammar for the resulting
textual CSM would include the productions \etexttt{$<$Expression$>$ ::= $<$BinaryExpression$>$} and \etexttt{$<$BinaryExpression$>$ ::=
$<$Expression$>$ $<$BinaryOperator$>$ $<$Expression$>$}, where associativity is not explicit. Left-associativity will be imposed by the
corresponding parser semantic action.

\subsubsection{The @Composition annotation}


The \an{@Composition} annotation allows the specification of the suitable order of evaluation of compositions represented in a CSM, a situation
that appears whenever the composite design pattern \cite{gof} is present in the ASM, no delimiters are employed to eliminate potential
ambiguities, and the composite contains several consecutive components of the same type of the composite. When such composites are nested,
different interpretations are possible unless we specify composition constraints in the CSM. This is the case of the typical shift-reduce
conflicts that appear in LR parsers when parsing nested if-then-else statements.

Hence, a specific constraint on element composition must be used to enforce a particular interpretation of such nested compositions in the
ASM-CSM mapping. ModelCC supports the following options for composition constraints:

\begin{itemize}

\item
\an{UNDEFINED}, when no composition constraints are defined and potential ambiguities are taken into account.

\item
\an{EAGER}, when the matching of constituent elements is performed as soon as possible. This corresponds
to the typical interpretation of nested if-then-else statements in programming languages, where the else clause is attached to the innermost if
statement.

\item
\an{LAZY}, when the matching of constituent elements is deferred as much as possible. Then, a rightmost derivation is obtained; i.e. when an
element might accompany any of two nested language constructs, it is associated to the outermost one.

\item
\an{EXPLICIT}, when no composition constraints are defined and any ambiguities should be resolved with the help of delimiters.

\end{itemize}


Composition order constraints are enforced by defining precedences for the productions in the grammar of the resulting textual CSM. Establishing
such precedences is possible in most parsing tools, including all yacc \cite{yacc} derivatives and Fence \cite{Quesada2011b}. When
composition is eager, shift operations will precede reduce operations. In contrast, when composition is lazy, reduce operations will have
precedence over shift operations. Finally, when the composition order must be explicit in the CSM, the use of delimiters will determine whether
shift or reduce operations are performed on a case by case basis.

\begin{figure}[tb!]
\centering
\includegraphics[scale=1]{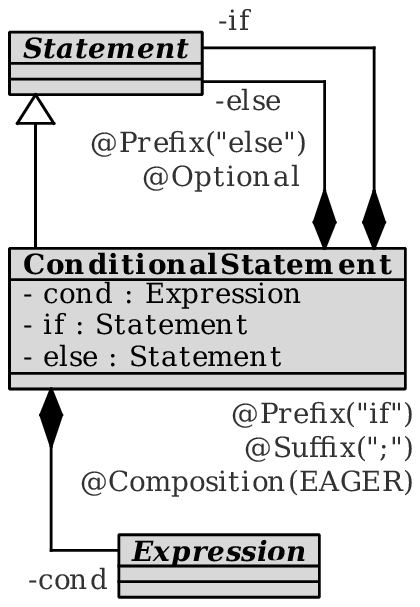}
\caption{Composition constraint example.}
\label{fig:conditional3}
\end{figure}

For example, the model in Figure \ref{fig:conditional3} represents typical if-then-else statements. In this case, the optional else
\emph{Statement} of the eager \emph{ConditionalStatement} will always match the innermost if statement when such statements are nested, e.g. in
``if E1 if E2 S1 else S2'', the else clause will correspond to the E2 if statement. The grammar for the resulting CSM will include the following
productions: \etexttt{$<$ConditionalStatement$>$ ::= ``if'' $<$Expression$>$ $<$Statement$>$ ``;'' $|$ ``if'' $<$Expression$>$ $<$Statement$>$
``else'' $<$Statement$>$ ``;''}. The parser will enforce the precedence of the second alternative over the first one, so that else clauses are
parsed as usual.

\subsubsection{The @Priority annotation}


The \an{@Priority} annotation allows the specification of precedences among language elements for eliminating ambiguities in textual CSMs.


ModelCC implements two mechanisms to specify priority constraints in the ASM-CSM mapping:

\begin{itemize}

\item
A relative one, where precedence relationships are established between particular language elements (a \emph{precedes} declaration indicates
which language elements have lower priority than the current element).

\item
An absolute one, where a numeric priority \emph{value} determines the priority level for each language element (the lower the value, the higher
the priority)

\end{itemize}

Unless specified otherwise, all language elements have the same priority. Precedences established among basic language elements must be managed
at the lexical analysis level. In ModelCC, the Lamb scanning algorithm \cite{Quesada2011a} enforces those lexical precedences. The Fence parsing algorithm
\cite{Quesada2011b} manages all the remaining precedences that can be established among the non-basic language elements that appear in
concatenation, selection, and repetition constructs within the CSM.


\begin{table*}[!htb]
\begin{center}

\setlength{\tabcolsep}{5pt}
\begin{tabular}{ l  l  l } \hline

Constraints on... & Annotation & Function \\ \hline

\multirow{2}{*}{Patterns}
& @Pattern & Pattern matching definition of basic language elements. \\
& @Value & Field where the recognized input element will be stored. \\ \hline

\multirow{3}{*}{Delimiters}
& @Prefix & Element prefix(es). \\
& @Suffix & Element suffix(es). \\
& @Separator & Element separator(s). \\ \hline

\multirow{3}{*}{Cardinality}
& @Optional & Optional elements.\\
& @Minimum & Minimum element multiplicity.\\
& @Maximum & Maximum element multiplicity.\\ \hline

\multirow{3}{50pt}{Evaluation order}
& @Associativity & Element associativity (e.g. left-to-right). \\
& @Composition & Eager or lazy composition for nested composites. \\
& @Priority & Element precedence level/relationships. \\ \hline
\end{tabular}

\end{center}
\caption{Summary of the metadata annotations supported by ModelCC.} \label{fig:tablesummary}
\end{table*}


\begin{figure}[tb!]
\centering
\includegraphics[scale=1]{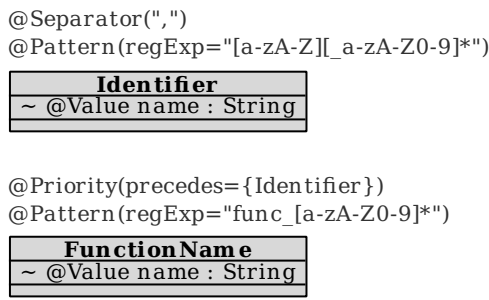}
\caption{Relative priority resolved by the lexical analyzer.} \label{fig:identifier3}
\end{figure}

\begin{figure}[tb!]
\centering
\includegraphics[scale=1]{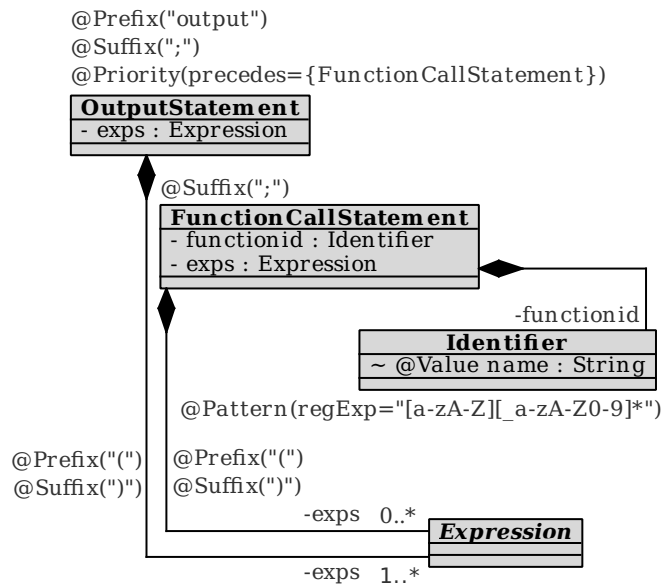}
\caption{Relative priority resolved by the syntactic analyzer.} \label{fig:outputfunction}
\end{figure}

For example, the model in Figure \ref{fig:identifier3} establishes a (fictitious) relative priority constraint between function names and
identifiers: a \emph{FunctionName} will always precede an \emph{Identifier}. In case a string like ``func\_power'' is found in the input string,
it will be recognized as a \emph{FunctionName}, but not as an \emph{Identifier}. Since the constraint is defined over basic language elements,
the lexical analyzer generated by ModelCC will be responsible for identifying the right element in the input string.

Figure \ref{fig:outputfunction} shows another example. In this case, the model enforces relative priority constraints between composite language
elements that will be resolver by the parser generated by ModelCC. Here, output statements precede function calls so that a string like
``output(3+5,4+1);'' will be recognized as an \emph{OutputStatement} but not as an \emph{FunctionCallStatement}, even when ``output'' would be a
perfectly valid indentifier. The ModelCC lexer, which supports lexical ambiguities, will consider ``output'' as an \emph{Identifier} and also as
a delimiter for output statements (i.e. its prefix keyword in the CSM).

\subsection{Summary of ModelCC ASM-CSM Constraints}

Table \ref{fig:tablesummary} summarizes the set of constraints supported by ModelCC for establishing ASM-CSM mappings between abstract syntax
models and their concrete representation in textual CSMs:

\begin{itemize}


\item
Pattern specification constraints are employed to specify the textual representation of basic language elements in the CSM. They define the
token types recognized by the lexer and indicate where the recognized tokens will be stored in the ASM.

\item
Delimiter constraints determine the prefixes, suffixes, and separators that will be used to mark the boundaries of language elements in the CSM.
They can be used for eliminating language ambiguities or just as syntactic sugar in text-based CSMs.

\item
Cardinality constraints restrict the multiplicity of repetitions and determine the optionality of language elements. The CSM must consider such
constraints for defining the grammar that defines the language recognized by the generated parser.

\item
Finally, evaluation order constraints allow the explicit resolution of different kinds of lexical and syntactic ambiguities that might appear in
the ASM-CSM mapping.

\end{itemize}

\section{A Working Example} \label{sec:example}

In this section, we compare how an interpreter for arithmetic expressions can be implemented using conventional tools and how it can be
implemented using the model-driven approach using ModelCC. Albeit the calculator example in this section is necessarily simplistic, it already
provides some hints on the potential benefits model-driven language specification can bring to more challenging endeavors.

First, we will outline the features we wish to include in our calculator language. Later, we will describe how an interpreter for this language is built
using two of the most established tools in use by language designers: lex \& yacc on the one hand, ANTLR on the other. Finally, we will
implement the same language processor using ModelCC by defining an abstract syntax model. This ASM will be annotated to specify the required
ASM-CSM mapping and it will also include the necessary logic for evaluating arithmetic expressions. This example will let us compare ModelCC
against conventional parser generators and it will be used for discussing the potential advantages provided by our model-driven language
specification approach.

\subsection{Language Description}

Our calculator will employ classical arithmetic expressions in infix notation. The language will feature the following capabilities:

\begin{itemize}

\item Unary operators: +, and -.
\item Binary operators: +, -, *, and /, being - and / left-associative.
\item Operator priorities: * and / precede + and -.
\item Parenthesized expressions.
\item Integer and floating-point literals.
\end{itemize}


\subsection{Conventional Implementation}

Using conventional tools, the language designer would start by specifying the grammar defining the calculator language in a BNF-like notation.
The BNF grammar shown in in Figure \ref{fig:calctrad} meets the requirements of our simple calculator, albeit it is not yet suitable for being
used with existing parser generators, since they impose specific constraints on the format of the grammar depending on the parsing algorithms
they employ.

\begin{figure*}[htp]
\begin{smallverbatim}
<Expression> ::= <ParenthesizedExpression>
               | <BinaryExpression>
               | <UnaryExpression>
               | <LiteralExpression>

<ParenthesizedExpression> ::= '(' <Expression> ')'

<BinaryExpression> ::= <Expression> <BinaryOperator> <Expression>

<UnaryExpression> ::= <UnaryOperator> <Expression>

<LiteralExpression> ::= <RealLiteral>
                      | <IntegerLiteral>

<BinaryOperator> ::= '+' | '-' | '/' | '*'

<UnaryOperator> ::= '+' | '-'

<RealLiteral> ::= <IntegerLiteral> '.' | <IntegerLiteral> '.' <IntegerLiteral>

<IntegerLiteral> ::= <Digit> <IntegerLiteral> | <Digit>

<Digit> ::= '0' | '1' |'2' | '3' | '4' | '5' | '6' | '7' | '8' | '9'

\end{smallverbatim}
\caption{BNF grammar for the arithmetic expression language.} \label{fig:calctrad}
\end{figure*}

\subsubsection{Lex \& yacc implementation}

When using lex \& yacc, the language designer converts the BNF grammar into a grammar suitable for LR parsing. A suitable
lex/yacc implementation defining the arithmetic expression grammar is shown in Figure \ref{fig:calclexyacc}.

\begin{figure*}[htp]
\begin{smallverbatim}
// Lex specification ['calc.lex']

%{
#include "y.tab.h"
extern YYSTYPE yylval;
%}
%%
[0-9]+\.[0.9]*  return RealLiteral;
[0-9]+          return IntegerLiteral;
\+|\-           return UnaryOrPriority2BinaryOperator;
\/|\*           return Priority1BinaryOperator;
\(              return LeftParenthesis;
\)              return RightParenthesis;
.               ;
%%


// Yacc specification ['calc.yacc']

%left UnaryOrPriority2BinaryOperator
%left Priority1BinaryOperator
%token IntegerLiteral RealLiteral LeftParenthesis RightParenthesis
%start Expression
%%
Expression : RealLiteral
           | IntegerLiteral
           | LeftParenthesis Expression RightParenthesis
           | UnaryOrPriority2BinaryOperator Expression
           | Expression UnaryOrPriority2BinaryOperator Expression
           | Expression Priority1BinaryOperator Expression
           ;
%%
#include "lex.yy.c"
int main(int argc,char *argv[]) { yyparse(); }
int yyerror(char *s) { printf("%s",s); }
\end{smallverbatim}
\caption{lex \& yacc specification of the arithmetic expression language.} \label{fig:calclexyacc}
\end{figure*}

Since lex does not support lexical ambiguities, the \emph{UnaryOperator} and \emph{BinaryOperator} nonterminals from the BNF grammar in
Figure \ref{fig:calctrad} have to be refactored in order to avoid the ambiguities introduced by the use of + and - both as unary and binary
operators. A typical solution consists of creating the \emph{UnaryOrPriority2BinaryOperator} token type for representing them and then adjusting
the grammar accordingly. This token will act as an \emph{UnaryOperator} in \emph{UnaryExpression}s, and as a \emph{BinaryOperator} in
\emph{BinaryExpression}s.

A similar solution is necessary for distinguishing different operator priorities, hence different token types are defined for each precedence
level in the language, even though they perform the same role from a conceptual point of view. The order in which they are declared in the
yacc specification determines their relative priorities (please, note that these declarations are also employed to define operator
associativity).

Unfortunately, the requirement to resolve ambiguities by refactoring the grammar defining the language involves the introduction of a certain
degree of duplication in the language specification: separate token types in the lexer and multiple parallel production rules in the parser.

Once all ambiguities have been resolved, the language designer completes the lex \& yacc introducing semantic actions to perform
the necessary operations. In this case, albeit somewhat verbose in C syntax, the implementation of an arithmetic expression evaluator is
relatively straightforward using the yacc \$ notation, as shown in Figure \ref{fig:calcimlexyacc}. In our particular implementation of
the calculator interpreter, carriage returns are employed to output results, hence our use of the ancillary \emph{Line} token type and
\emph{LineReturn} nonterminal symbol.

\begin{figure*}[htp]
\begin{smallverbatim}
// Lex specification ['calc.lex']

%{
#include "string.h"
#include "y.tab.h"
extern YYSTYPE yylval;
%}
%%
[0-9]+\.[0.9]*  { yylval.value = atof(yytext); return RealLiteral; }
[0-9]+          { yylval.value = (double)atoi(yytext); return IntegerLiteral; }
\+|\-           {
                  if (yytext[0] == '+') yylval.operator = PLUSORADDITION;
                  else /* yytext[0] == '-' */ yylval.operator = MINUSORSUBSTRACTION;
                  return UnaryOrPriority2BinaryOperator;
                }
\/|\*           {
                  if (yytext[0] == '*') yylval.operator = MULTIPLICATION;
                  else /* yytext[0] == '/' */ yylval.operator = DIVISION;
                  return Priority1BinaryOperator;
                }
\(              { return LeftParenthesis; }
\)              { return RightParenthesis; }
\n              { return LineReturn; }
.               ;
%%


// Yacc specification ['calc.yacc']

%left UnaryOrPriority2BinaryOperator
%left Priority1BinaryOperator
%token IntegerLiteral RealLiteral LeftParenthesis RightParenthesis
%start Line
%{
#include <stdio.h>
#define YYSTYPE attributes
typedef enum { PLUSORADDITION, MINUSORSUBSTRACTION, MULTIPLICATION, DIVISION } optype;
typedef struct {
  optype operator;
  double value;
} attributes;
%}
%%
Expression : RealLiteral       { $$.value = $1.value; }
           | IntegerLiteral    { $$.value = $1.value; }
           | LeftParenthesis Expression RightParenthesis  { $$.value = $2.value; }
           | UnaryOrPriority2BinaryOperator Expression
             {
               if ($1.operator == PLUSORADDITION) $$.value = $2.value;
               else /* $1.operator == MINUSORSUBSTRACTION */ $$.value = -$2.value;
             }
           | Expression UnaryOrPriority2BinaryOperator Expression
             {
               if ($2.operator == PLUSORADDITION) $$.value = $1.value+$3.value;
               else /* $2.operator == MINUSORSUBSTRACTION */ $$.value = $1.value-$3.value;
             }
           | Expression Priority1BinaryOperator Expression
             {
               if ($2.operator == MULTIPLICATION) $$.value = $1.value*$3.value;
               else /* $2.operator == DIVISION */ $$.value = $1.value/$3.value;
             }
           ;
Line       : Expression LineReturn { printf("%f\n",$1.value); } ;
%%
#include "lex.yy.c"
int main(int argc,char *argv[]) { yyparse(); }
int yyerror(char *s) { printf("%s",s); }
\end{smallverbatim}
\caption{Complete lex \& yacc implementation of the arithmetic expression interpreter.} \label{fig:calcimlexyacc}
\end{figure*}

\subsubsection{ANTLR implementation}

When using ANTLR, the language designer converts the BNF grammar into a grammar suitable for LL parsing. An ANTLR specification of
our arithmetic expression language is shown in Figure \ref{fig:calcantlr}.

\begin{figure*}[h!tp]
\begin{smallverbatim}
grammar ExpressionEvaluator;

expression1 : expression2 ( '+' expression1 | '-' expression1 )* ;

expression2 : expression3 ( '*' expression2 | '/' expression2 )* ;

expression3 : '(' expression1 ')'
            | '+' expression1
            | '-' expression1
            | INTEGERLITERAL
            | FLOATLITERAL
            ;

INTEGERLITERAL : '0'..'9'+ ;

FLOATLITERAL : ('0'..'9')+ '.' ('0'..'9')* ;

NEWLINE : '\r'? '\n' ;
\end{smallverbatim}
\caption{ANTLR specification of the arithmetic expression language.} \label{fig:calcantlr}
\end{figure*}

Since ANTLR provides no mechanism for the declarative specification of token precedences, such precedences have to be incorporated into
the grammar. The usual solution involves the creation of different nonterminal symbols in the grammar, so that productions corresponding to the
same precedence levels are grouped together. The productions with \emph{expression1} and \emph{expression2} in their left-hand side were
introduced with this purpose in our calculator grammar.

Likewise, since ANTLR generates a LL(*) parser, which does not support left-recursion, left-recursive grammar productions in the grammar
shown in Figure \ref{fig:calctrad} have to be refactored. In our example, a simple solution involves the introduction of the \emph{expression3}
nonterminal, which in conjunction with the aforementioned \emph{expression1} and \emph{expression2} nonterminals, eliminates left-recursion from
our grammar.

Once the grammar is adjusted to satisfy the constraints imposed by the ANTLR parser generator, the language designer can define the
semantic actions needed to implement our arithmetic expression interpreter. The resulting ANTLR implementation is shown in Figure
\ref{fig:calcimantlr}. The streamlined syntax of the scannerless ANTLR parser generator makes this implementation significantly more
concise than the equivalent lex \& yacc implementation. However, the constraints imposed by the underlying parsing algorithm
forces explicit changes on the language grammar (cf. BNF grammar in Figure \ref{fig:calctrad}).

\begin{figure*}[tp]
\begin{smallverbatim}
grammar ExpressionEvaluator;

expression1 returns [double value]
            : e=expression2 {$value = $e.value;}
              ( '+' e2=expression1 {$value += $e2.value;}
              | '-' e2=expression1 {$value -= $e2.value;}
              )*
            ;

expression2 returns [double value]
            : e=expression3 {$value = $e.value;}
              ( '*' e2=expression2 {$value *= $e2.value;}
              | '/' e2=expression2 {$value -= $e2.value;}
              )*
            ;

expression3 returns [double value]
            : '(' e=expression1 ')' {$value = $e.value;}
            | '+' e=expression1 {$value = $e.value;}
            | '-' e=expression1 {$value = -$e.value;}
            | i=INTEGERLITERAL {$value = (double)Integer.parseInt($i.text);}
            | f=FLOATLITERAL {$value = Double.parseDouble($f.text);}
            ;

INTEGERLITERAL : '0'..'9'+ ;

FLOATLITERAL : ('0'..'9')+ '.' ('0'..'9')* ;

NEWLINE : '\r'? '\n' ;
\end{smallverbatim}
\caption{Complete ANTLR implementation of the arithmetic expression interpreter.}
\label{fig:calcimantlr}
\end{figure*}

\subsection{ModelCC Implementation}

\begin{figure*}[htb]
\centering
\includegraphics[scale=1]{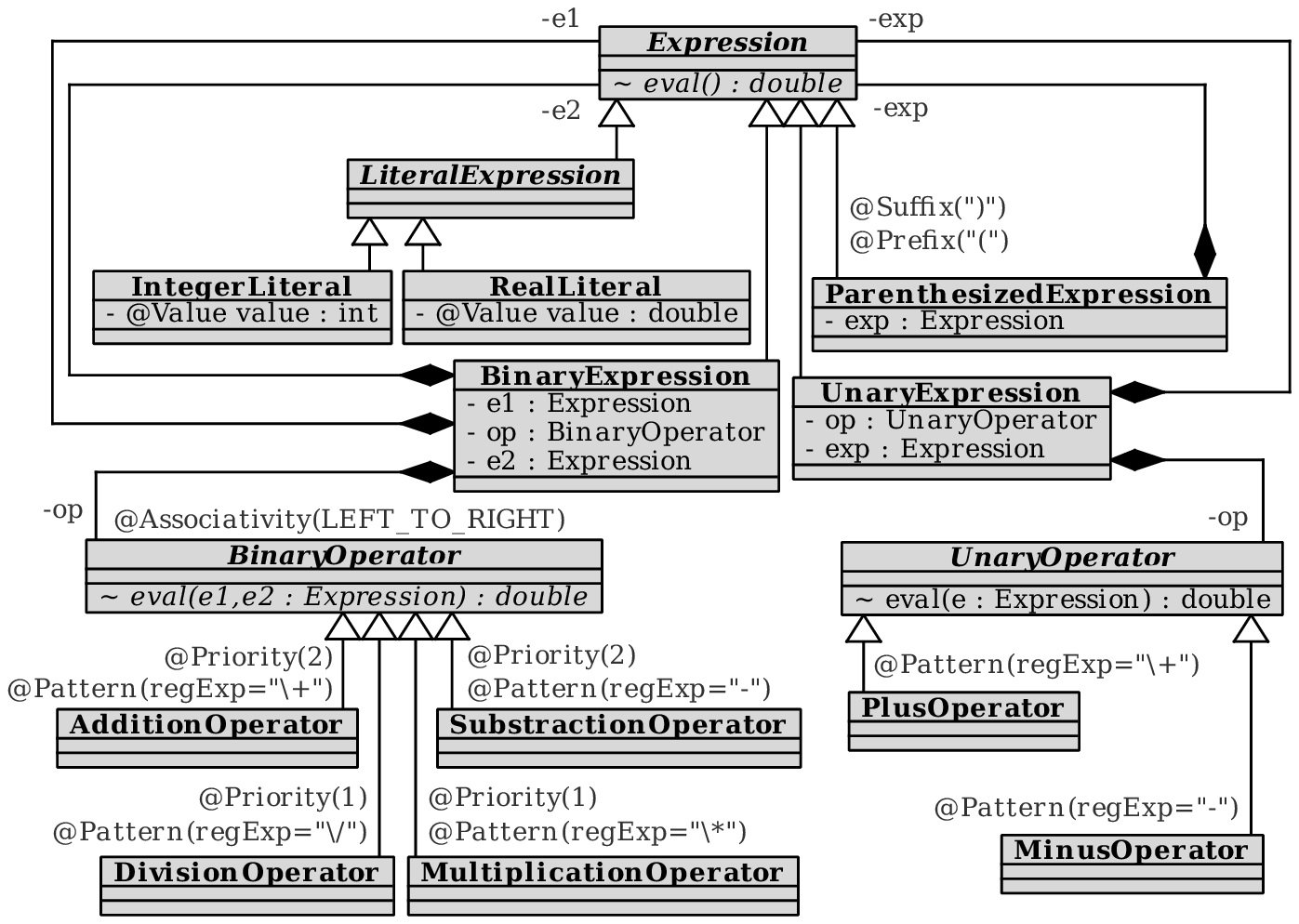}
\caption{ModelCC specification of the arithmetic expression language.} \label{fig:calcmodelcc}
\end{figure*}

When following a model-based language specification approach, the language designer starts by elaborating an abstract syntax model, which will
later be mapped to a concrete syntax model by imposing constraints on the abstract syntax model. These constraints can also be specified as
metadata annotations on the abstract syntax model and the resulting annotated model can be processed by automated tools, such as ModelCC, to
generate the corresponding lexers and parsers. Annotated models can be represented graphically, as the UML diagram in Figure
\ref{fig:calcmodelcc}, or implemented using conventional programming languages, as the Java implementation listed in Figure
\ref{fig:calcimmodelcc}.

\begin{figure*}[htp]
\begin{smallverbatim}
public abstract class Expression implements IModel {
  public abstract double eval();
}
\end{smallverbatim}
\sepv
\begin{smallverbatim}
@Prefix("\\(") @Suffix("\\)")
public class ParenthesizedExpression extends Expression implements IModel {
  Expression e;
  @Override public double eval() { return e.eval(); }
}
\end{smallverbatim}
\sepv
\begin{smallverbatim}
public abstract class LiteralExpression extends Expression implements IModel {
}
\end{smallverbatim}
\sepv
\begin{smallverbatim}
public class UnaryExpression extends Expression implements IModel {
  UnaryOperator op;
  Expression e;
  @Override public double eval() { return op.eval(e); }
}
\end{smallverbatim}
\sepv
\begin{smallverbatim}
public class BinaryExpression extends Expression implements IModel {
  Expression e1;
  BinaryOperator op;
  Expression e2;
  @Override public double eval() { return op.eval(e1,e2); }
}
\end{smallverbatim}
\sepv
\begin{smallverbatim}
public class IntegerLiteral extends LiteralExpression implements IModel {
  @Value int value;
  @Override public double eval() { return (double)value; }
}
\end{smallverbatim}
\sepv
\begin{smallverbatim}
public class RealLiteral extends LiteralExpression implements IModel {
  @Value double value;
  @Override public double eval() { return value; }
}
\end{smallverbatim}
\sepv
\begin{smallverbatim}
public abstract class UnaryOperator implements IModel {
  public abstract double eval(Expression e);
}
\end{smallverbatim}
\sepv
\begin{smallverbatim}
@Pattern(regExp="\\+")
public class PlusOperator extends UnaryOperator implements IModel {
  @Override public double eval(Expression e) { return e.eval(); }
}
\end{smallverbatim}
\sepv
\begin{smallverbatim}
@Pattern(regExp="-")
public class MinusOperator extends UnaryOperator implements IModel {
  @Override public double eval(Expression e) { return -e.eval(); }
}
\end{smallverbatim}
\sepv
\begin{smallverbatim}
@Associativity(AssociativityType.LEFT_TO_RIGHT)
public abstract class BinaryOperator implements IModel {
  public abstract double eval(Expression e1,Expression e2);
}
\end{smallverbatim}
\sepv
\begin{smallverbatim}
@Priority(value=2) @Pattern(regExp="\\+")
public class AdditionOperator extends BinaryOperator implements IModel {
  @Override public double eval(Expression e1,Expression e2) { return e1.eval()+e2.eval(); }
}
\end{smallverbatim}
\sepv
\begin{smallverbatim}
@Priority(value=2) @Pattern(regExp="-")
public class SubstractionOperator extends BinaryOperator implements IModel {
  @Override public double eval(Expression e1,Expression e2) { return e1.eval()-e2.eval(); }
}
\end{smallverbatim}
\sepv
\begin{smallverbatim}
@Priority(value=1) @Pattern(regExp="\\*")
public class MultiplicationOperator extends BinaryOperator implements IModel {
  @Override public double eval(Expression e1,Expression e2) { return e1.eval()*e2.eval(); }
}
\end{smallverbatim}
\sepv
\begin{smallverbatim}
@Priority(value=1) @Pattern(regExp="\\/")
public class DivisionOperator extends BinaryOperator implements IModel {
  @Override public double eval(Expression e1,Expression e2) { return e1.eval()/e2.eval(); }
}
\end{smallverbatim}
\caption{Complete Java implementation of the arithmetic expression interpreter using ModelCC: A set of Java classes define the language ASM,
metadata annotations specify the desired ASM-CSM mapping, and object methods implement arithmetic expression evaluation.}
\label{fig:calcimmodelcc}
\end{figure*}

In the current version of ModelCC, annotated models representing the ASM and a particular ASM-CSM mapping are used to generate Lamb
lexers \cite{Quesada2011a} and Fence parsers \cite{Quesada2011b}, albeit traditional LL and LR parsers might also be generated whenever
the ASM-CSM mapping constraints make LL and LR parsing feasible.

Since the abstract syntax model in ModelCC is not constrained by the vagaries of particular parsing algorithms, the language design process can
be focused on its conceptual design, without having to introduce artifacts in the design just to satisfy the demands of particular tools:

\begin{itemize}

\item
As we saw in the lex \& yacc example, conventional tools, unless they are scannerless, force the creation of artificial token
types in order to avoid lexical ambiguities, which leads to duplicate grammar production rules and semantic actions in the language
specification. As in any other software development project, duplication hinders the evolution of languages and affects the maintainability of
language processors. ModelCC, even though it is not scannerless, supports lexical ambiguities and each basic language element is defined as a
separate and independent entity, even when their pattern specification are in conflict. Therefore, duplication in the language model does not
have to be included to deal with lexical ambiguities: token type definitions do not have to be adjusted, duplicate syntactic constructs rules
will not appear in the language model, and, as a consequence, semantic actions do not have to be duplicated either.

\item
As we also saw both in the lex \& yacc calculator and in the ANTLR solution to the same problem, established parser
generators require modifications to the language grammar specification in order to comply with parsing constraints, let it be the elimination of
left-recursion for LL parsers or the introduction of new nonterminals to restructure the language specification so that the desired precedence
relationships are fulfilled. In the model-driven language specification approach, the left-recursion problem disappears since it is something
the underlying tool can easily deal with in a fully automated way when an abstract syntax model is converted into a concrete syntax model.
Moreover, the declarative specification of constraints, such as the evaluation order constraints in Section \ref{sec:evaluation-order}, is
orthogonal to the abstract syntax model that defines the language. Those constraints determine the ASM-CSM mapping and, since ModelCC takes
charge of everything in that conversion process, the language designer does not have to modify the abstract syntax model just because a given
parser generator might prefer its input in a particular format. This is the main benefit that results from raising your abstraction level in
model-based language specification.

\item
When changes in the language specification are necessary, as it is often the case when a software system is successful, the traditional language
designer will have to propagate changes throughout the entire language processing tool chain, often introducing significant changes and making
profound restructurings in the working code base. The changes can be time-consuming, quite tedious, and extremely error-prone. In contrast,
modifications are more easily done when a model-driven language specification approach is followed. Any modifications in a language will affect
either to the abstract syntax model, when new capabilities are incorporated into a language, or to the constraints that define the ASM-CSM
mapping, whenever syntactic details change or new CSMs are devised for the same ASM. In either case, the more time-consuming, tedious, and
error-prone modifications are automated by ModelCC, whereas the language designer can focus his efforts on the essential part of the required
changes.

\item
Finally, traditional parser generators typically mix semantic actions with the syntactic details of the language specification. This approach,
which is justified when performance is the top concern, might lead to poorly-designed hard-to-test systems when not done with extreme care.
Moreover, when different applications or tools employ the same language, any changes to the syntax of that language have to be replicated in all
the applications and tools that use the language. The maintenance of several versions of the same language specification in parallel might also
lead to severe maintenance problems. In contrast, the separation of concerns provided by ModelCC, as separate ASM and ASM-CSM mappings, promotes
a more elegant design for language processing systems. By decoupling language specification from language processing and providing a conceptual
model for the language, different applications and tools can now use the same language without having duplicate language specifications. A
similar result could be hand-crafted using traditional parser generators (i.e. making their implicit conceptual model explicit and working on
that explicit model), but ModelCC automates this part of the process.

\end{itemize}

In summary, while traditional language processing tools provide different mechanisms for resolving ambiguities and implementing language
constraints, the solutions they provide typically interfere with the conceptual modeling of languages: relatively minor syntactic details might
significantly affect the structure of the whole language specification. Model-driven language specification, as exemplified by ModelCC, provides
a cleaner separation of concerns: the abstract syntax model is kept separate from its incarnation in concrete syntax models, thereby separating
the specification of abstractions in the ASM from the particularities of their textual representation in CSMs.

\section{Conclusions and Future Work} \label{sec:conclusionsfuturework}

In this paper, we have introduced ModelCC, a model-based tool for language specification. ModelCC lets language designers create explicit models
of the concepts a language represents, i.e. the abstract syntax model (ASM) of the language. Then, that abstract syntax can be represented in
textual or graphical form, using the concrete syntax defined by a concrete syntax model (CSM). ModelCC automates the ASM-CSM mapping by means of
metadata annotations on the ASM, which let ModelCC act as a model-based parser generator.

ModelCC is not bound to particular scanning and parsing techniques, so that language designers do not have to tweak their models to comply with the
constraints imposed by particular parsing algorithms. ModelCC abstracts away many details traditional language processing tools have to deal
with. It cleanly separates language specification from language processing. Given the proper ASM-CSM mapping definition, ModelCC-generated
parsers are able to automatically instantiate the ASM given an input string representing the ASM in a concrete syntax.

Apart from being able to deal with ambiguous languages, ModelCC also allows the declarative resolution of any language ambiguities by means of
constraints defined over the ASM. The current version of ModelCC also supports lexical ambiguities and custom pattern matching classes. A
fully-functional version of ModelCC is available at http://www.modelcc.org.

The proposed model-driven language specification approach promotes the domain-driven design of language processors. Its model-driven philosophy
supports language evolution by improving the maintainability of languages processing system. It also facilitates the reuse of language
specifications across product lines and different applications, eliminating the duplication required by conventional tools and improving the
modularity of the resulting systems.


In the future, ModelCC will incorporate a wider variety of parsing techniques and it will be able to automatically determine the most efficient
parsing algorithm that is able to parse a particular language (the current version employs the Fence parsing algorithm
\cite{Quesada2011b} on top of the Lamb scanning algorithm \cite{Quesada2011a}).

ModelCC will also be extended to support multiple concrete syntax models for the same abstract syntax model.

We plan to study the possibilities tools such as ModelCC open up in different application domains, including traditional language processing
systems (compilers and interpreters) \cite{Aho2006}, domain-specific languages (DSLs) \cite{Fowler2010,Hudak1996,Mernik2005} and language
workbenches \cite{language-workbenches}, model-driven software development (MDSD) tools \cite{Schmidt2006}\cite{mdsd-ideal}, natural language
processing \cite{Jurafsky2009} in restricted domains, model induction, text mining applications, data integration, and information extraction.


\bibliographystyle{plain}
\bibliography{doc}

\end{document}